\documentclass[
12pt,
reprint,
showpacs,preprintnumbers,
showkeys,
amsmath,amssymb,
aps,
prb,
]{revtex4-1}

\usepackage{pdfpages}
\makeatletter
\AtBeginDocument{\let\LS@rot\@undefined}
\makeatother
\usepackage{graphicx}
\usepackage{dcolumn}
\usepackage{bm}
\usepackage{hyperref}
\usepackage{nicefrac}
\usepackage{color}
\usepackage{braket}
\usepackage{multirow}
\usepackage{verbatim}
\usepackage{bbold}
\usepackage[caption=false]{subfig}

\newcommand{\LIA}{\lambda_{\rm I}^{\rm A}}
\newcommand{\LIB}{\lambda_{\rm I}^{\rm B}}
\newcommand{\LI}{\lambda_{\rm I}}
\newcommand{\LR}{\lambda_{\rm R}}

\newcommand{\K}{\rm{K}}
\newcommand{\Kp}{\rm{K}^\prime}

\begin{document}

\title{Protected pseudohelical edge states in proximity graphene ribbons and flakes}

\author{\surname{Tobias} Frank}
\email[Emails to: ]{tobias1.frank@physik.uni-regensburg.de}
\author{\surname{Petra} H\"{o}gl}
\author{\surname{Martin} Gmitra}
\author{\surname{Denis} Kochan}
\author{\surname{Jaroslav} Fabian}

\affiliation{%
 Institute for Theoretical Physics, University of Regensburg,\\
 93040 Regensburg, Germany
 }%

\date{\today}

\begin{abstract}
We investigate topological properties of models that describe graphene on realistic substrates which induce proximity spin-orbit coupling in graphene.  A $\mathbb{Z}_2$  phase diagram is calculated for the parameter space of (generally different) intrinsic spin-orbit coupling on the two graphene sublattices, in the presence of Rashba coupling. The most fascinating case is that of staggered intrinsic spin-orbit coupling which, despite being topologically trivial, $\mathbb{Z}_2 = 0$, does exhibit edge states protected against time-reversal scattering for zigzag ribbons as wide as micrometers. We call these states pseudohelical as their helicity is locked to the sublattice. The spin character and robustness of the pseudohelical modes is best exhibited on a finite flake, which shows that the edge states have zero $g$-factor, carry a finite spin current in the crossection of the flake, and exhibit spin-flip reflectionless tunneling at the armchair edges.  
\end{abstract}

\maketitle

Graphene is an exciting material to investigate electrical transport\cite{Neto2009}, but it also has remarkable spin properties that make it useful for spintronics applications\cite{Zutic2004, Fabian2007}. One outstanding issue in graphene spintronics \cite{han2014} is the enhancement of spin-orbit coupling (SOC) which is only about 10 $\mu$eV in pristine graphene\cite{Gmitra2009}. Perhaps the principal drive for enhancing SOC is our desire to realize in graphene topological effects such as the quantum spin Hall state (QSHS)\cite{Kane2005, Kane2005a}, anomalous quantum Hall effect\cite{Qiao2010, Yang2011}, or topological superconductivity\cite{Ren2016}.

\begin{figure}[htp]
     \includegraphics{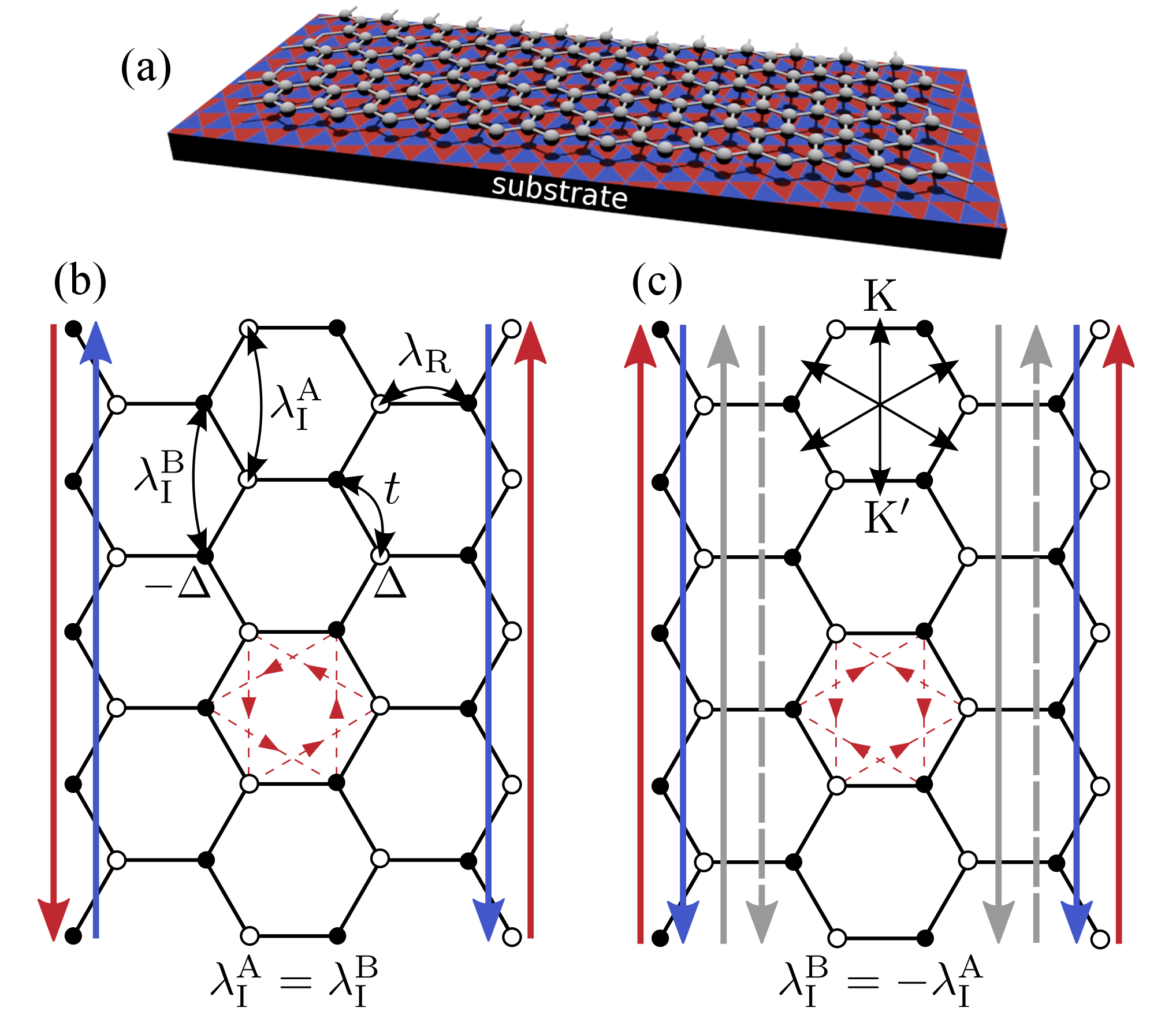} 
     \caption{\label{fig:schemes} (Color online) Schematics of proximity induced properties in graphene. Figure (a) shows graphene placed on a sublattice symmetry breaking substrate.
    Information in Figs.~(b) and (c) encoded in black shows bulk graphene related information. Symbols colored in red (blue) denote spin-up (spin-down) characteristics. Sublattice A is represented as empty dots, sublattice B as filled dots. Fig.~(b) shows the hopping parameters used in our tight-binding model. Dashed red lines encode the next-nearest neighbor (spin-dependent) spin preserving hoppings, for the uniform case of $\lambda_\mathrm{I}^\mathrm{A}
= \lambda_\mathrm{I}^\mathrm{B}$, within a hexagon. Helical edge states and their velocity directions are indicated by long arrows. Figure (c) shows the reciprocal $\mathrm{K}$ and $\mathrm{K}^\prime$ directions with respect to the real space lattice. The next-nearest neighbor hoppings in a hexagon are shown for our case of staggered
intrinsic spin-orbit coupling, $\lambda_\mathrm{I}^\mathrm{A}
= - \lambda_\mathrm{I}^\mathrm{B}$. Solid (dashed) gray arrows indicate valley edge states located in the $\kappa=1 (-1)$ valley. Red and blue arrows show the pseudohelical states carrying a finite spin current along the ribbon.} 
\end{figure}

The most promising way to induce large and uniform SOC in graphene is via proximity effects which allow  graphene to inherit some properties from the substrate. There is, however, a trade-off. Often the substrates break the sublattice (pseudospin) symmetry, which has two important effects. First, an orbital gap opens due to the staggered potential, and, second, intrinsic spin-orbit coupling acquires a staggered term. In the original model of Kane and Mele\cite{Kane2005}, the intrinsic coupling on A and B sublattices is the same. In proximity graphene, they are in general different, such as graphene on copper\cite{Frank2016}. The most extreme case is graphene on transition metal dichalcogenides (TMDCs), schematically depicted in Fig.~\ref{fig:schemes}(a). By these substrates spin-valley locking is induced in graphene, manifested in the appearance of the valley Zeeman coupling---opposite (in sign) intrinsic SOCs in the sublattices\cite{Gmitra2015, Gmitra2016, Wang2015}. Rashba coupling is also induced. 
Valley Zeeman effect is proposed to be detected as a giant spin lifetime anisotropy\cite{Cummings2017}.  
There are already experiments on graphene on TMDCs\cite{Avsar2014, Lu2014, Larentis2014, Wang2015, Wang2016, Yang2016, Omar2017, Volkl2017}. Weak localization \cite{Wang2015, Wang2016, Yang2016, Volkl2017} and spin transport measurements\cite{Avsar2014, Omar2017, Dankert2016} confirm the proximity induced SOC in graphene in the range of \mbox{1--10~meV}. Density functional theory calculations predict SOC of about 1 meV\cite{Gmitra2015, Gmitra2016, Wang2015, Yang2016, Kaloni2014}. In the extreme case of strong SOC, as in graphene on WSe$_2$, inverted band structure arises \cite{Gmitra2016, Wang2015,  Alsharari2016} indicating the possibility of topological edge states. At the moment there is no consistent picture. 
It was reported that the inverted structure is topologically nontrivial \cite{Wang2015}, 
which appears in line with the claim of helical edge modes \cite{Gmitra2016}, but 
inconsistent with the statement that the system has $\mathbb{Z}_2 = 0$ and there are no topologically protected edge states\cite{Wang2015}. 
We aim to provide a unified picture of, on one hand, the topological nature of  proximity models, and, on the other hand, the existence and character of protected edge states. We introduce a modified Haldane model\cite{Haldane1988} with staggered intrinsic spin-orbit coupling to illustrate how the edge states appear in models with spin-valley locking. There are in general two pairs of edge states formed at each edge. This makes the bulk model trivial, which we confirm by a map of the $\mathbb{Z}_2$ invariant, an overdue 
reference needed to classify  proximity graphene on a specific substrate. 
Can protected edge states arise in such a trivial system? Yes, and the key is to gap
out unwanted (valley) pair of states.  This is effortlessly realized in ribbons (of micron
sizes), as we show. The remaining pair is protected against time reversal scattering, just like the QSHS. 
But unlike helical states of the QSHS, our edge states are pseudohelical, being spin up at one zigzag edge, and spin down at the other. These states are connected by reflectionless
spin-flip scattering at the armchair edges in a flake. Pseudohelical states carry spin current, and have zero $g$-factor. 

The electronic structure of a bipartite hexagonal lattice with broken sublattice and broken horizontal reflection symmetry, such as graphene on a substrate, can be described by the $C_{3v}$-symmetric Hamiltonian\cite{Gmitra2013, Gmitra2016, Kochan2017},
\begin{eqnarray}\label{eq:ham_tb}
    \mathcal{H} &=&
    \sum_{\left<i,j\right>} \,t\, c_{is}^\dagger c^{\phantom\dagger}_{js}+
    \sum_i \,\Delta \, \xi_i \,c_{is}^\dagger c^{\phantom\dagger}_{is} \nonumber \\
    &&+\frac{2i}{3}\sum_{\left<i,j\right>}c_{is}^\dagger c^{\phantom\dagger}_{js'}\left[\LR\left(\hat{\mathbf{s}}\times \mathbf{d}_{ij}\right)_z\right]_{ss^\prime}\nonumber\\
    &&+\frac{i}{3}\sum_{\left<\left<i,j\right>\right>}c_{is}^\dagger c^{\phantom\dagger}_{js^\prime} \left[\frac{\lambda_{\rm I}^{i}}{\sqrt{3}}\nu_{ij}\hat{\mathbf{s}}_z \right]_{ss^\prime}\,.
\end{eqnarray}
The hopping terms are schematically depicted in Fig.~\ref{fig:schemes}(b). The nearest neighbor hopping $t$ occurs between sites $i$ and $j$, preserving spin $s$. The staggered potential $\Delta$ has signs $\xi_i=1$ and $-1$,  for sublattice A and B, respectively. Horizontal reflection symmetry is broken by the Rashba SOC $\LR$ which mixes states of opposite spins and sublattices. The unit vector $\mathbf{d}_{ij}$ points from site $j$ to site $i$ and $\hat{\mathbf{s}}$ is the vector of spin Pauli matrices. The last term, the intrinsic SOC, is a next-nearest neighbor hopping. It couples same spins and depends on clockwise ($\nu_{ij}= - 1$) or counterclockwise ($\nu_{ij}= 1$) paths along a hexagonal ring from site $j$ to site $i$. This Hamiltonian distinguishes intrinsic SOC at different sublattices $\LI^i$, where $i$ stands for A or B. This is the principal extension of the models  introduced earlier by Haldane \cite{Haldane1988} and by Kane and Mele\cite{Kane2005} (in fact, already McClure and Yafet\cite{McClure1962} introduced intrinsic SOC for graphene). This extension makes the models experimentally
relevant, while introducing  new physics. 

Fourier transformation and linearization of Hamiltonian (\ref{eq:ham_tb}) around the $\K$ and $\Kp$ points addressed by the valley index $\kappa = \pm 1$, respectively, results in the sum of the following Hamiltonians\cite{Kochan2017, Gmitra2015}
\begin{eqnarray}
    \mathcal{H}_{\rm k} &=& \hbar v_{\rm F} (\kappa k_x \sigma_x - k_y \sigma_y) s_0,\label{eq:graphene}\\
    \mathcal{H}_{\rm \Delta} &=& \Delta\,\sigma_z s_0,\label{eq:staggered}\\
    \mathcal{H}_{\rm R} &=& \LR(-\kappa\sigma_x s_y+\sigma_y s_x), \\
    \mathcal{H}_{\rm I} &=& \frac{1}{2} \left [\LIA  (\sigma_z+\sigma_0) + \LIB(\sigma_z-\sigma_0)\right ] \kappa s_z,\label{eq:intrinsic}
\end{eqnarray}
corresponding to the order of terms in Eq.~(\ref{eq:ham_tb}).
The Fermi velocity $v_{\rm F}$ is expressed as $\sqrt{3}at/2\hbar$ with lattice constant $a$. The sublattice (pseudospin) degrees of freedom are described by Pauli matrices $\sigma$.
Following Kane and Mele \cite{Kane2005}, we use in this work for numerical examples values of $t=1$, $\Delta=0.1\,t$, $\LR=0.075\,t$, and $\LIA,|\LIB| = \sqrt{27}\cdot0.06\,t$ if not indicated differently. In reality we expect
weaker couplings from proximity effects\cite{Gmitra2016}, but here our goal is to demonstrate qualitative features of the models.  We will also comment on what one can expect quantitatively in real samples.

To illustrate the physics of our model, let us first look only at spin up (spinless) electrons and choose the two opposite limits $\LIA = \LIB = \LI$ as the {\it uniform}, and $\LIB = -\LIA = \LI$ as the {\it staggered} intrinsic SOC model cases. The corresponding Hamiltonians
are, 
\begin{align}
    \label{eq:intrinsic_km}\mathcal{H}_{\rm I}^{\rm uniform} =& \LI \sigma_z \kappa,\\
    \label{eq:intrinsic_akm}\mathcal{H}_{\rm I}^{\rm staggered} =& \LI \sigma_0 \kappa.
\end{align}
\begin{figure}
    \includegraphics[width=0.99\columnwidth]{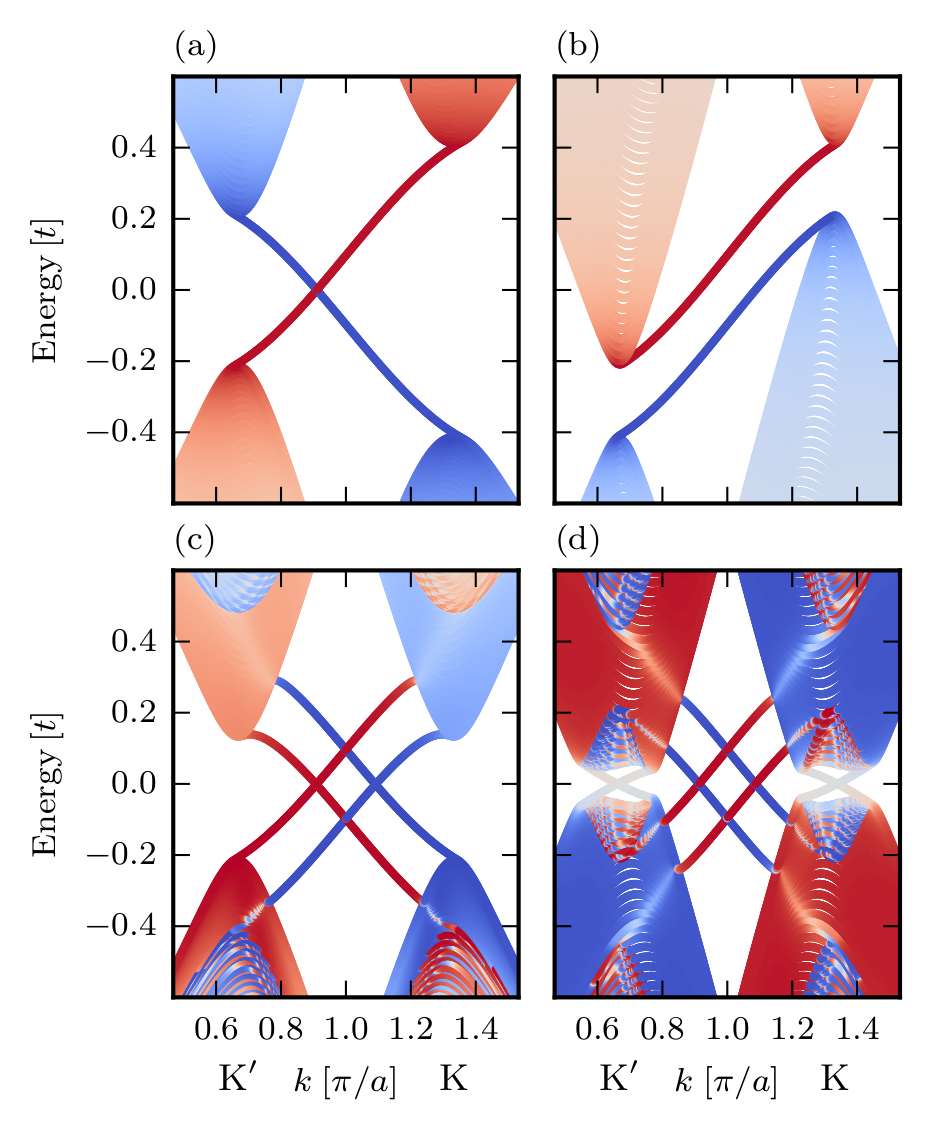}
    \caption{\label{fig:zigzag} (Color online) Spectrum of a zigzag ribbon of a width of 100 graphene unit cells. The color code in (a) and (b) for the spinless case denotes the sublattice expectation value, red for sublattice A, and blue for sublattice B. The spectrum of the spinful case in  (c) and (d) is color coded with the spin expectation value, red for spin-up, and blue for spin-down. Left column shows the uniform case, $\LIA = \LIB$, right column staggered case with 
strong spin-valley locking, $\LIB = -\LIA$.}
\end{figure}
The energy spectrum of a zigzag ribbon for spin-up electrons is plotted in  Figs.~\ref{fig:zigzag} (a) and (b). 
The two valleys with bulk-like subbands are well visible. Between valley maxima and minima, edge modes appear due to the chiral nature of graphene\cite{Haldane1988}. While for the uniform case the two edge states have opposite
velocities, in our case of spin-valley locking, the edge states have the same velocity, producing spin-up current in the ground state. 

The spectra in Figs.~\ref{fig:zigzag} (a) and (b) can be understood from simple considerations. For $\K$ electrons, the phase of the Bloch wave function on sublattice A/B rotates (increases by $2\pi/3$) counterclockwise/clockwise. For $\Kp$ electrons this behavior is reversed. The staggered potential leads to the following pseudospin-valley state:
$(v\K, c\K; v \Kp, c\Kp) = (B, A; B, A)$; here $c$ and $v$ label the conduction and valence bands.   
We now add intrinsic SOC, which can be viewed as an action of a vector potential 
(Peierls phase), whose rotation within the sublattices is sketched in Figs.~\ref{fig:schemes}(b) and (c). 
If the Bloch phase rotation has the same sense as the rotation of the vector potential, the energy of the state
increases. If the rotations are opposite, the energy decreases. (This is analogous to a system with an orbital
momentum in a magnetic field.) 

In the uniform case, the vector potential rotates counterclockwise [Fig.~\ref{fig:schemes}(b)] so that at 
$\K$ electrons in sublattice A gain energy and B lose. The opposite is true at $\Kp$, with the result seen in Fig.~\ref{fig:zigzag}(a). This establishes the connection to Eq.~(\ref{eq:intrinsic_km}), which is a valley-pseudospin Zeeman coupling. 
Once the effective magnetic field $\LI$ overcomes the staggered potential $\Delta$, the sublattice occupation 
becomes (B, A; A, B), flipping A and B at $\Kp$, and a chiral state that crosses the gap develops, shown in Fig.~\ref{fig:zigzag}(a). This is the well known case of a Chern insulator\cite{Haldane1988}.

In the case of staggered intrinsic SOC regime, the Peierls field acts on each sublattice equally in each valley [see Fig.~\ref{fig:schemes}(c)]. The energy levels shift in opposite directions in the two valleys, and the sublattice expectation values remain (B, A; B, A); see Fig.~\ref{fig:zigzag}(b). The Hamiltonian in Eq.~(\ref{eq:intrinsic_akm}) represents
a valley Zeeman coupling. If $\LI \geq \Delta$ the system becomes metallic, as the conduction band in the $\Kp$ point has lower energy than the valence band in the $\K$ point. Nevertheless, there are isolated propagating states, which connect states of same sublattice expectation value from the different valleys.

Let us now reinstate both spins into the picture. The complete spectra for zigzag ribbons are obtained by 
mirroring the spectra in Figs.~\ref{fig:zigzag}(a) and (b) around the time reversal invariant point $\pi/a$.
If we also introduce Rashba SOC, we get additional spin mixing. The results are shown in 
Figs.~\ref{fig:zigzag}(c) and (d). In the uniform case, the resulting band structure is additive, leading to two pairs of helical edge states, a manifestation of the QSHS\cite{Kane2005}. The only effect of Rashba SOC is the mixing of spins in the bulk bands, most apparent in the conduction bands.

In the staggered case, Fig.~\ref{fig:zigzag}(d), there are also (what appears to be) helical edge modes present,
as in the QSHS, which would have an energetic overlap with the bulk states if Rashba SOC is absent. 
Contrary to the QSHS, the edge states with same spin on different edges travel along the {\it same} direction, see Fig.~\ref{fig:schemes}(c), leading to a net spin current. The effect of Rashba SOC leads to the opening of a bulk gap in the valleys due to the different spin expectation values of valence and conduction bands. This gap is an inverted one, as parts of the former valence bands are now higher in energy than parts of the former conduction bands, which is called mass inversion\cite{Alsharari2016}. Inside this Rashba gap two new edge states appear in each valley, with 
quenched spins. Each valley contributes one mode per edge with opposite velocities on the distinct boundaries [see Fig.~\ref{fig:schemes}(c)]. Having both valley-centered and helical states, we term this quantum valley spin Hall state (QVSHS).

\begin{figure}
 \includegraphics[width=0.98\columnwidth]{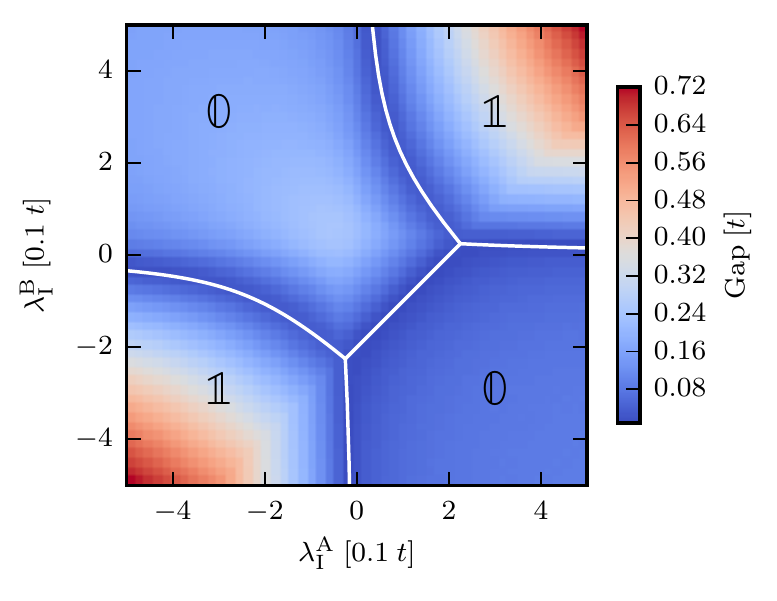}
 \caption{\label{fig:phase_space} (Color online) $\mathbb{Z}_2$ phase space and bulk gap landscape of graphene in the $\LIA-\LIB$ plane. Color denotes the size of the gap in graphene. Solid lines are analytical expressions for a (global) bulk graphene gap closing, which separate trivial ($\mathbb{0}$) and non-trivial ($\mathbb{1}$) phases from each other as indicated. Orbital parameters and Rashba SOC are the same as in the text.}
\end{figure}

In general the symmetries of a bulk Hamiltonian can be used to classify its topological properties\cite{Ryu2010}. The Hamiltonian in Eq.~(\ref{eq:ham_tb}) possesses time-reversal symmetry, has broken particle-hole and sublattice symmetries, which puts it into the class of AII Hamiltonians\cite{Ryu2010}. In two dimensions this leads to the possibility of a $\mathbb{Z}_2$ classification, which for our set of models is shown in Fig.~\ref{fig:phase_space}, in the 
space of the two sublattice intrinsic SOC parameters. This map shows four distinct regions separated by gap closings, where one can expect a change in topology. $\mathbb{Z}_2$ invariants are calculated numerically \cite{Gresch2017} (see suppl. mat.). The QSHS regions in the upper right and lower left corners exhibit nontrivial 
topologies, while the QVSHS, focused at $\LIA = -\LIB$ diagonal, are $\mathbb{Z}_2$ trivial. Bulk band structure schemes 
of representing this phase diagram are in the supplemental material.

We find the staggered cases to have a trivial $\mathbb{Z}_2$ invariant, as stated already in Ref.~\onlinecite{Yang2016} for $\LIA = -\LIB$. In addition, we find that unlike zigzag ribbons, armchair ones have no edge states. The valley Chern number in the staggered case is 1 (see suppl. mat.) as found also in Ref.~\onlinecite{Alsharari2016}. This Chern number characterizes the states that occur inside the valley and confirms the existence of one conducting channel per edge and valley. We note that our system regarding the valley centered states is very similar to bilayer graphene subject to a perpendicular electric field which shows a quantum valley Hall state (QVHS)\cite{Qiao2011}. This system represents twice a copy of our one with a valley Chern number of 2 due to spin degeneracy, showing an absence of states in armchair ribbons as well. This absence is due to intervalley (short-range) scattering as $\K$ and $\Kp$ are mapped onto each other in the armchair geometry.

\begin{figure}[htp]
     \includegraphics{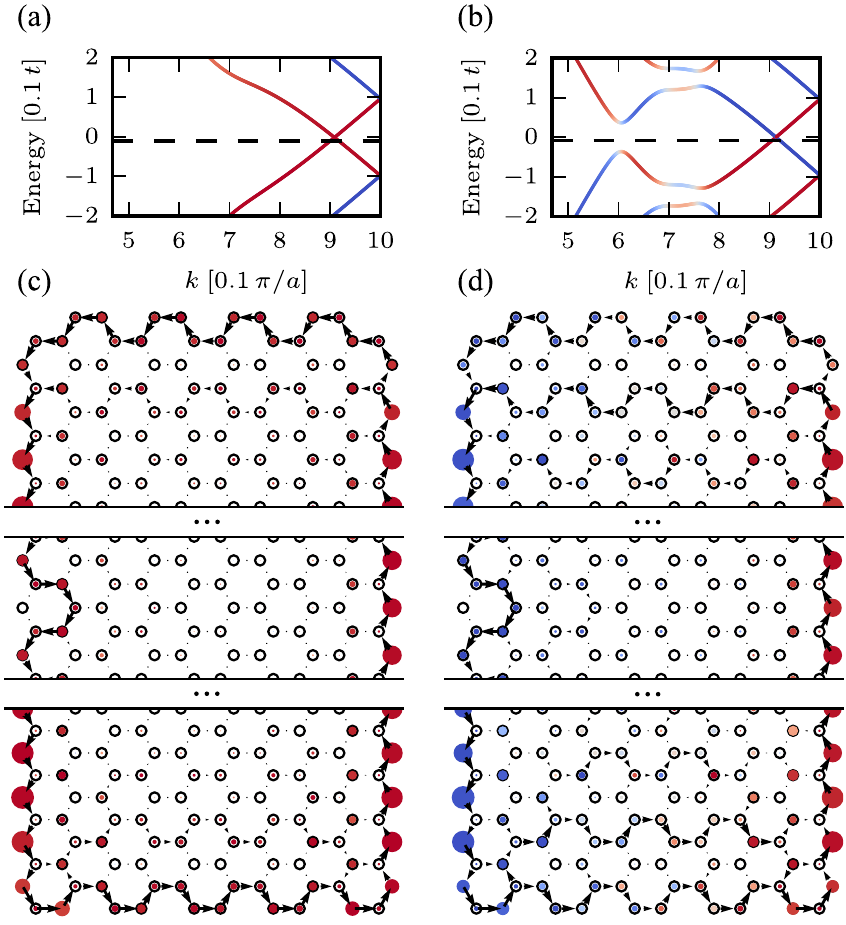} 
     \caption{\label{fig:flakes}(Color online) Finite sized zigzag ribbons and flakes of width of ten zigzag unit cells. Left column is for the case of $\LIA = \LIB$ and right column for $\LIB = -\LIA$. Figures (a) and (b) shows the band structure of an infinite zigzag ribbon over half of the Brillouin zone with spin expectation values as color code (up in red, down in blue). Figures (c) and (d) show finite flakes of length of 100 zigzag cells and properties of a state that lies at energy indicated by dashed lines in Figs.~(a) and (b), respectively. Empty dots denotes the lattice, full dots indicate the site expectation value color coded for spin polarization and black arrows show particle bond currents. An orbital in the middle panels of Figs.~(c) and (d) have been removed acting as a short range scatterer. Flakes have been cut due to size constraints.}
\end{figure}

Crucial for our further analysis is the localization behavior of the edge states. To get the localization length we fit
$|\Psi(y)|^2 \propto \exp(-y/\lambda)$ (see suppl. mat.), where $y$ is measured from the edge. We find that the spin-polarized edge states decay very fast, over half a unit cell ($\lambda \approx 0.4\,a$), whereas the valley states have a much longer localization length ($\lambda \approx 9\,a$). This indicates that for narrow ribbons valley states should be gapped due to hybridization. A comparison of the band structures for zigzag ribbons of width of ten unit cells for uniform and staggered case is shown in Figs.~\ref{fig:flakes}(a) and (b), respectively. Indeed, the valley states exhibit a gap in the QVSHS in Fig.~\ref{fig:flakes}(b).

The reason for the larger decay length of the valley states is that they are spectrally close to the bulk states, see Fig.~\ref{fig:zigzag}(d). We find the relation $6 t/E_\mathrm{g} > w/a$ between the value of inverted gap $E_{\mathrm{g}}$ and the zigzag ribbon width $w$, for which valley edge states gap out due to finite size quantization effects (see suppl. mat.). For realistic gaps\cite{Gmitra2016} (in the order of meVs) valley states should not affect the edge physics of ribbons more narrow than 10000 unit cells (about $2.6\,\mu$m), wide enough for experimental
investigations.

With the valley states gapped out, we are left with a single pair of spin-polarized states at each edge inside the
gap. What are these states and how do they compare to the helical modes of the QSHS? In particular, since the
spin-up modes head in one direction along the two edges, how do the states meet in a finite flake?   
To clarify this question, we calculated finite graphene flakes taking states from within the gap as shown in Fig.~\ref{fig:flakes}(a) and (b). To simulate short-range scattering we removed one orbital from the left zigzag edge (setting the onsite energy there to $10^8\,t$). Additionally we calculated spin and site expectation values as well as probability bond currents\cite{Boykin2010}. In the QSHS, Fig.~\ref{fig:flakes}(c), we find as expected a true helical edge state flowing along the boundary, avoiding the short range scatterer and preserving its spin along $z$. 
The time reversal partner of this state has the opposite chirality and opposite spin polarization (not shown).

Edge states appearing in the finite-size gap of QVSHS are shown in Fig.~\ref{fig:flakes}(d). They have several
fascinating features. (a) The probability bond current navigates around the short range scatterer and \textit{does not scatter back}. The reason is that there is only the time-reversal partner, $T\psi$ of the edge state $\psi$ at this energy and, as for topologically protected states, backscattering is forbidden as long as the impurity $V$ is nonmagnetic and scattering is elastic (mathematically, $\langle \psi| V |T\psi \rangle = 0$).  (b) Spin polarization is opposite on the
two edges, which are formed by the two sublattices. This is why we call these states {\it pseudohelical}---with
pseudo describing either the pseudospin-spin locking or ``not-really-helical'' character of the states. Net spin current
flows in this state along the zigzag direction. Also, we explicitly checked that the out-of-plane $g$-factor of the
pseudohelical states is zero, as expected since, although they are locally spin polarized, globally the pseudohelical states are spinless. 
(c) Finally, also at odds with true helical states which exist along armchair, pseudohelical states exhibit  
reflectionless tunneling through the armchair boundary. The tunneling is assisted by Rashba SOC providing
the necessary spin flip. The coupling of the edge states is done via the continuing of velocities, not via the spin
character. 

To conclude, we need to enrich the parameter space of spin-orbit Hamiltonians when dealing with proximity
graphene. Novel models include the extreme spin-valley locking in which intrinsic SOC is opposite on the two 
sublattices. We provide the full map of the $\mathbb{Z}_2$ invariant for this extended class of parameters
and show that the spin-valley locking models are $\mathbb{Z}_2$ trivial. Nevertheless, we prove that robust 
protection against back scattering can be induced in finite ribbons of micron sizes, by gapping out unwanted
states and leaving only what we call pseudohelical states in the gap which have fascinating properties.
These findings are important for graphene on substrates such as TMDCs, especially with the ability of atomically precise growth of zigzag ribbons\cite{Ruffieux2016}.  Of interest is also the general idea that we can stabilize edge (and perhaps surface) states against elastic time-reversal scattering by gapping out an otherwise coexisting class of other states.  

\begin{acknowledgments}
This work was supported by the DFG SFB Grant No. 689 and GRK Grant No. 1570, and the International Doctorate Program Topological Insulators of the Elite Network of Bavaria. This project has received funding from the European Union's Horizon 2020 research and innovation programme under grant agreement No. 696656. The authors gratefully acknowledge the Gauss Centre for Supercomputing e.V. (www.gauss-centre.eu) for funding this project by providing computing time on the GCS Supercomputer SuperMUC at Leibniz Supercomputing Centre (LRZ, www.lrz.de).
\end{acknowledgments}

\bibliography{paper}

\onecolumngrid
\newpage



\section*{Supplementary material}

# Protected pseudohelical edge states in proximity graphene ribbons and flakes

Tobias Frank, Petra Högl, Martin Gmitra, Denis Kochan, and Jaroslav Fabian*

*Institute for Theoretical Physics, University of Regensburg,*
*93040 Regensburg, Germany*
(Dated: July 7, 2017)

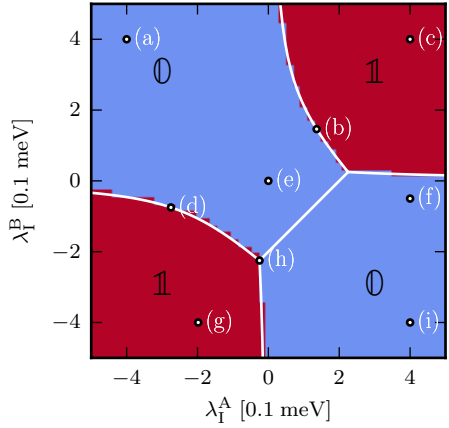

FIG. S1. (Color online) Calculated $\mathbb{Z}_2$ invariant as a function of intrinsic spin-orbit coupling. Blue areas and red areas have trivial (0) and nontrivial (1) $\mathbb{Z}_2$ invariants, respectively. White lines denote analytical gap closing conditions at the K point. Labeled dots (a) to (i) show positions of selected parameters whose band structures are discussed in section S2.

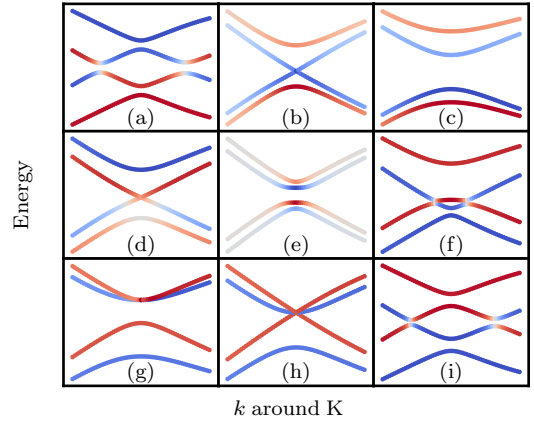

FIG. S2. (Color online) Graphene bulk band topologies around the K point for selected intrinsic SOC parameters. Panel labeling (a) to (i) is in one-to-one correspondence with intrinsic SOC parameters displayed in Fig. S1 labelled by (a) to (i). Color encodes spin-expectation values, red and blue stand correspondingly for spin-up and spin-down projections.

## S1. $\mathbb{Z}_2$ INVARIANT CALCULATION

Our Hamiltonian, given in Eq. (1), possesses time-reversal symmetry, has broken particle-hole as well as broken sublattice symmetry, which puts it into the class of AII Hamiltonians[1]. In two dimensions we can classify the Hamiltonian in terms of a $\mathbb{Z}_2$ invariant. To calculate explicitly the $\mathbb{Z}_2$ invariant, we used the code Z2PACK[2]. With the help of tracking Wannier charge centers of the occupied bands over half of the Brillouin zone, one can determine the $\mathbb{Z}_2$ invariant of the system. Results for the phase space with varying intrinsic parameters are shown in Fig. S1, where we keep parameters of $t = 1$, $\Delta = 0.1\,t$, $\lambda_R = 0.075\,t$. These parameters and $\lambda_I^A$, $|\lambda_I^B| = \sqrt{27} \cdot 0.06\,t$ are used in this supplemental material if not indicated differently, as in the main text. The convergence threshold for the Wannier positions was set to POSTOL $= 10^{-4}$ in order to get converged results. We sampled the phase space by a $51 \times 51$ grid, which closely follows the expected behavior of the $\mathbb{Z}_2$ invariant being 1 in the quantum spin Hall state (QSHS) and 0 in the regions which can be reached from the QSHS only by closing the graphene bulk gap.

## S2. BAND TOPOLOGIES

The Hamiltonian, Eq. (1), presented in the main text is very rich. Therefore we selected special points from the phase space in Fig. S1 to illustrate which bulk band topologies one can expect and display them in Fig. S2. Representative figures for the staggered case are Figs. S2(a) and (i), Figs. S2(c) and (g) show the case of quantum spin Hall systems. Figure S2(b) shows the closing of the gap when going from the uniform SOC to the staggered SOC case. Figure S2(e) is taken from a point, where the intrinsic spin-orbit coupling (SOC) is zero and shows an example where also the zigzag ribbons are gapped and no edge states are present (not shown). Special point Fig. S2(d) shows the case of a gap closing with maximal mixture of spins in the valence bands. In Fig. S2(h) we show the case of a triple degeneracy, which lies at a point where three gap closing lines meet each other. A more generic case for the inverted mass regime is shown in Fig. S2(f).

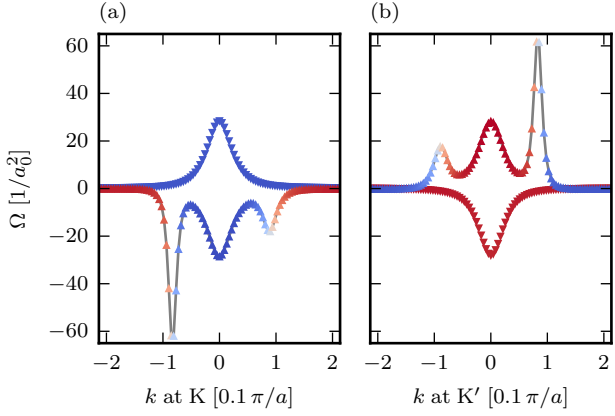

FIG. S3. (Color online) Berry curvature for the two low energy states of bulk graphene in the staggered case. Berry curvature is shown at K in (a) and K' in (b). Downward pointing triangles denote the band lower in energy, upward pointing triangles indicate the band higher in energy. Color encodes spin-expectation value, where red and blue stand for spin-up, and spin-down, respectively.

## S3. BERRY CURVATURE

To characterize the origin of the edge states in the staggered SOC case, we carried out the analysis of Berry curvatures,

$$\Omega_n(\mathbf{k}) = -2\mathrm{Im} \sum_{n' \neq n} \frac{\langle u_n^{\mathbf{k}} | \hat{v}_x^{\mathbf{k}} | u_{n'}^{\mathbf{k}} \rangle \langle u_{n'}^{\mathbf{k}} | \hat{v}_y^{\mathbf{k}} | u_n^{\mathbf{k}} \rangle}{(\omega_{n'}^{\mathbf{k}} - \omega_n^{\mathbf{k}})^2}, \quad (S2)$$

as they can be used to calculate Chern numbers and tell about the existence of conducting edge channels. $\hat{v}_i$ is the velocity operator in direction $i$, $\omega_n^{\mathbf{k}} = E_n^{\mathbf{k}}/\hbar$ with $E_n^{\mathbf{k}}$ being the energy eigenvalues of the Hamiltonian.

Berry curvatures for the staggered case of occupied bands at K/K' points are shown in Fig. S3, corresponding to a bulk band structure qualitatively the same as in Fig. S2(i). The asymmetry stems from trigonal warping effects. The central peaks at the K/K' points are responsible for the pseudohelical modes, where for example in the QSHS without Rashba SOC a spin Chern number can be defined[3]. At the gaps, which are opened due to the Rashba SOC at the Fermi energy [compare Fig. S2(i)], side peaks in the Berry curvature appear. This is understandable, as the Berry curvature is usually high at small-gapped anticrossings, see Eq. (S2).

The valley Chern number is defined as[3]

$$\mathcal{C}_v = (\mathcal{C}_K - \mathcal{C}_{K'})/2, \quad (S3)$$

with

$$\mathcal{C}_K = \frac{1}{2\pi} \sum_n^{occ} \int_{\mathbf{k} \in \mathrm{BZ}/2} d\mathbf{k} \ \Omega_n^z(\mathbf{k}). \quad (S3)$$

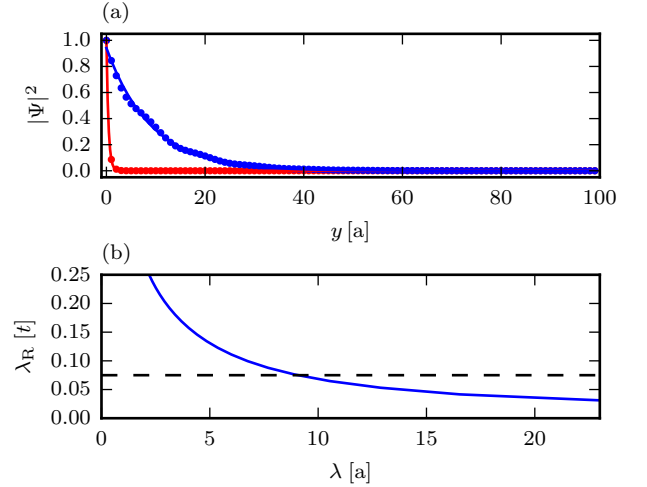

FIG. S4. (Color online) Localization behavior of pseudohelical and valley centered edge states of zigzag ribbon of width 100 unit cells with staggered parameters. Blue and red dots in (a) represent the probability of finding the valley centered and pseudohelical state in the $y$th unit cell, solid lines are the respective fits. Wave functions were renormalized to match in magnitude at the edge of the ribbon. (b) shows the localization length $\lambda$ of the valley centered state versus the Rashba parameter. The Rashba parameter used here is indicated as dashed line.

The integration limit extends over half of the Brillouin zone belonging to the triangle around the K/K' point.

We find the valley Chern number to be 1, with $\mathcal{C}_{K/K'} = \pm 1$ as found in Ref. 4 for the staggered case. We note that swapping of signs of $\lambda_I^A$ and $\lambda_I^B$ switches the signs of $\mathcal{C}_{K/K'}$ and leads to a valley Chern number of $-1$ as found in Ref. 4. The calculated valley Chern number is independent of the size of Rashba SOC, which we checked by varying it by $\pm 25\%$. In conclusion, the valley Chern number can serve as a fingerprint of the existence of valley states for a given parameter regime.

## S4. LOCALIZATION PROPERTIES

Edge states are expected to be exponentially localized at the boundaries of a ribbon of finite width. Wave functions of valley centered and pseudohelical states located at the left edge of a zigzag ribbon of width of 100 unit cells are shown in Fig. S4(a). The states were selected close to the respective crossing points in Fig. 2(d) from the lower energy edge state at momentum $k = 0.681\,\pi/a$ and $k = 0.917\,\pi/a$ for valley centered and pseudohelical states, respectively. We fit absolute squared wave functions obtained from our tight-binding calculations to an exponential function $|\Psi(y)|^2 \propto \exp(-y/\lambda)$, where $y$ is the transversal direction of the ribbon and $\lambda$ the decay length.

The valley centered states are found to have a decay

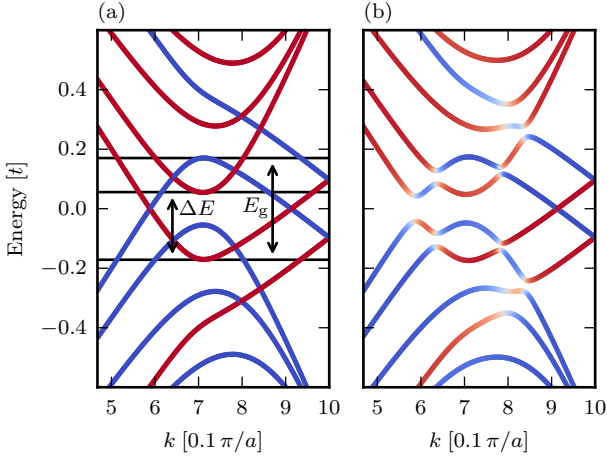

FIG. S5. (Color online) Finite size and Rashba coupling influence on zigzag ribbon band structures. Size of the ribbons is thirteen zigzag unit cells. (a) and (b) show band structures with and without Rashba SOC. Color encodes spin-expectation value, where red and blue stand for spin-up, and spin-down, respectively. Finite size level spacing $\Delta E$ and inverted gap size $E_\text{g}$ are indicated.

length of $\lambda \approx 9.0\,a$, whereas the pseudohelical states drop quickly with $\lambda \approx 0.4\,a$. The localization of the valley centered state depends strongly on the value of the Rashba coupling, see Fig. S4(b). The valley centered edge states live inside the Rashba gap. The larger the Rashba gap, the more the edge states are energetically decoupled from the bulk states, therefore the localization gets smaller with increasing Rashba SOC. This could have also potential for applications, as the Rashba coupling can be controlled by a perpendicular electric field.

## S5. LIMITATION BY FINITE SIZE AND GENERATION OF VALLEY CENTERED STATES

Besides the Rashba coupling, another important influence on the valley centered states stems from the finite width of the zigzag ribbons, which can be guessed from the localization behavior in Sec. S4. It can also be seen from the band structure of a narrow ribbon in Fig. S5. Part (a) shows how the interplay of intrinsic SOC and staggered potential creates gap inversion analogously to the one in main text Figure 2(d), but without Rashba SOC. One can recognize valence band maximum and conduction band minimum whose distance defines

the inverted gap $E_\text{g}$. From these valence and conduction extremal points the pseudohelical states originate and meet at the Brillouin zone boundary $k = \pi/a$. If Rashba SOC is added see Fig. S5(b), it splits spin-up and -down bands of different sublattices. The pseudohelical edge states are not affected as they are spatially decoupled.

The finite ribbon can be also viewed as a finite quantum well with hard wall boundaries, where bulk states are separated by a constant energy of $\Delta E$, as indicated in Fig. S5(a). We find this level spacing to be $\Delta E \approx (6/\sqrt{3})\hbar v_\text{F}/w$, which is proportional to the characteristic length scale in graphene divided by the width of the ribbon. The condition of not having the first subband cross the Fermi energy translates into

$$\Delta E > \frac{E_\text{g}}{2}, \qquad (S5)$$

as the inverted band gap does not depend on the width of the ribbon for not too small ribbon widths and assuming $\lambda_\text{R} \ll E_\text{g}$. Inserting $v_\text{F} = \sqrt{3}at/2\hbar$ into the equation one gets

$$\frac{w}{a} \lesssim 6\,\frac{t}{E_\text{g}}, \qquad (S5)$$

where the right hand side is the graphene tight-binding band width divided by the inverted gap. For the staggered case we get a critical value of $w \approx 14\,a$, inserting a gap of $E_\text{g} \approx 2\lambda_\text{I} - 2\Delta = 0.424\,t$. For more realistic gaps (e.g. WSe$_2$, Ref. 5) of $E_\text{g} = 1.32\,\text{meV} = 0.53 \cdot 10^{-3}\,t$ one would expect ribbons of width $10000\,a$ ($2.46\,\mu\text{m}$), where one can still expect only pseudohelical states to be present at the Fermi level. In this limit, to observe pseudohelical states one should reside within the Rashba gap to the left of the K point, see Fig. S5(b). This gap increases with the ribbon width, as observed in Ref. 5, with a power law dependence. In the limit of infinite ribbons this power law behavior would saturate at twice the Rashba parameter value.

So far we discussed finite size effects on the band structure of zigzag ribbons. To see how the valley states emerge from the bulk band structures we attached an animation. This video shows the same data as in Fig. S4 varying the width of the ribbon. With increasing width $\Delta E$ decreases due to finite size quantization and without Rashba coupling the states within the inverted gap become increasingly dense. Rashba coupling then acts among this densely sampled space and leaves back the valley states.

---


%
%
%

\pagestyle{empty}

\end{document}